\documentstyle[seceq,epsf,wrapfig,twoside]{ptptex}
\def\Journal#1#2#3#4{{#1}{\bf #2} (#4), #3}
\def\PTP{Prog.~Theor.~Phys.~}

\def\NPB{Nucl.~Phys. \bf{B}}
\def\PLB{Phys.~Lett. \bf{B}}

\def\PRL{Phys.~Rev.~Lett.~}
\def\PRD{Phys.~Rev. \bf{D}}
\def\ZPC{Z.~Phys. \bf{C}}

\def\PR{Phys.~Rev.~}
\def\AP{Ann.~Phys.~}
\notypesetlogo  
\title{%
Possible Classification of the Chiral Scalar $\sigma$-Nonet
}
\author{%
Muneyuki {\sc Ishida}
}
\inst{%
Department of Physics, 
Tokyo Institute of Technology, 
Tokyo 152-8551, Japan\\
}
\recdate{%
July 22 , 1998
}

\abst{%
We recently observed 
the iso-singlet scalar $\sigma$(600)-particle and 
the iso-doublet scalar $\kappa$(900)-particle,
through the re-analyses of $\pi\pi$- 
and $K\pi$-scattering phase shifts, respectively.
First, assuming 
the two iso-singlet states, the $\sigma$(600) and 
the established $f_0$(980), being the ideal mixing 
states $n\bar n$ and $s\bar s$, respectively, 
of a single scalar nonet,  
it is pointed out that 
the mass value of the iso-singlet flavor-octet state, obtained 
from the orthogonal transformation,
satisfies, together with the $\kappa$(900) and the 
iso-triplet scalar $a_0(980)$-particle,
 the Gell-Mann Okubo mass-formula.
Furthermore, it is argued that,
by investigating their properties of masses and widths,
this scalar $\sigma$-nonet, together with 
the pseudo-scalar $\pi$-nonet, realizes the 
linear representation of $SU(3)$ chiral symmetry.
}

\begin{document}
\maketitle
\setcounter{tocdepth}{4}

\section{Introduction\\
---{\it Observation of $\sigma$(600) and $\kappa$(900)}---}

In previous works by our group,
we have observed the scalar $q\bar q$-mesons with $I$=0 and $I$=1/2,
$\sigma (600)$\cite{rf:pipip,rf:further,rf:had97,rf:had97a} 
and $\kappa$(900),\cite{rf:piK,rf:had97,rf:had97a} respectively, 
by the re-analyses\footnote{
Other recent phase shift analyses also 
suggest the existence of light 
$\sigma$\cite{rf:achasov,rf:kamin,rf:torn,rf:hara} and 
$\kappa$\cite{rf:beve} particles.
}
 of the $\pi\pi$- and $K\pi$-scattering phase shifts.
The mass and width of the $\sigma$($\kappa$)-meson are
determined with the values 
of $m_\sigma =540- 675$ MeV\footnote{
In a previous work\cite{rf:further}
we estimated $m_\sigma$  
with the value of $585\pm 20$ MeV 
for the ``standard'' phase shift
of the CERN-M$\ddot{\rm u}$nich b-analysis.\cite{rf:CM}
} 
and $\Gamma_{\sigma\rightarrow\pi\pi} =385\pm 70$ MeV
($m_\kappa =
905\stackrel{+65}{\scriptstyle -30}$ MeV
and $\Gamma_{\kappa\rightarrow K\pi}=
545\stackrel{+235}{\scriptstyle -110}$ MeV).
Existence of these scalar mesons has been conventionally
neglected\cite{rf:Morgan,rf:morg,rf:pennington} for many years. 
The reason which led us to 
different results using the same data\cite{rf:CM,rf:LASS} 
are two-fold: 
Technically, we have applied a new $S$-matrix 
parametrization method, the interfering Breit-Wigner amplitude
(IA-method)\cite{rf:pipip,rf:further,rf:had97,rf:had97a,rf:piK} 
for the analyses,
where the amplitude is represented directly by 
the physically meaningful parameters,
masses and widths of resonances, while 
physically,
we have introduced a negative background phase 
$\delta_{
\rm BG}$\cite{rf:pipip,rf:further,rf:had97,rf:had97a,rf:piK,rf:had97b}
of a hard core type.\cite{rf:acore,rf:Ncore,rf:yazaki} 
This $\delta_{\rm BG}$ represents a very strong 
repulsive force between pions (or pion and kaon), which strongly
cancels\cite{rf:MY} the attractive force 
due to intermediate $\sigma$- (or $\kappa$-)
production. 
The obtained $\chi^2$value is greatly improved compared with that in the 
conventional analysis without the $\sigma$ (and $\kappa$) meson.
The physical reason\cite{rf:had97b} for missing the existence 
of these new scalar particles in the conventional phase shift analyses
is due to overlooking this cancellation mechanism, which is 
guaranteed by current algebra and PCAC.

In addition to these new particles, there are the other established
scalar mesons below 1 GeV: the iso-singlet $f_0(980)$ and the 
iso-triplet $a_0(980)$. 
The purpose of this paper is to investigate the possibility for  
classification of these scalar particles into a single scalar nonet,
and furthermore, to study 
whether it is possible to ascribe this scalar $\sigma$-nonet 
to the chiral partner of the pseudo-scalar
$\pi$-nonet in $SU(3)$ chiral symmetry.

In \S 2, the Gell-Mann Okubo (GMO) mass formula is shown
to be approximately satisfied for the octet members of this scalar 
$\sigma$-nonet,
where, in order to identify the octet members, the OZI-rule 
concerning the decay property
of $f_0(980)$ is assumed. 
In \S 3.1, by investigating the relations among their 
masses and widths, they are suggested to have the properties
of the scalar nonet predicted by the $SU(3)$ linear 
$\sigma$ model
(L$\sigma$M\cite{rf:LsM,rf:kramer,rf:BL,rf:CH,rf:gg}).
Being based on these results, in \S 3.2
the situations of the validity of OZI rule, which was 
assumed in \S 2, is examined somewhat quantitatively. 
Section 4 is devoted to concluding remarks.


\section{Scalar $\sigma$-nonet mass formula}

Experimentally,
$f_0(980)$ has a considerably small $\pi\pi$-width 
regardless of its large phase volume, while having a rather  
large $K\bar K$-width in spite of the fact that 
its mass is quite close
to the $K\bar K$-threshold.
Assuming the approximate validity of the OZI rule, 
this fact seems to suggest that 
$f_0(980)$ consists of almost 
pure $s\bar s$-component.
Here we simply assume that $\sigma$(600) and $f_0(980)$ are the 
ideal mixing\cite{rf:ideal} states
of a single scalar nonet and that the squared-mass matrix takes 
a diagonal form in these ideal bases.
The ideal-mixing states are related 
to the octet state $\sigma_8$ and the singlet state $\sigma_1$
through the orthogonal transformation:
\begin{eqnarray}
\left(
\begin{array}{c}
\sigma (600)\\
f_0(980)
\end{array}
\right)
=
\left(
\begin{array}{c}
\sigma_n\\
\sigma_s
\end{array}
\right)
 &=& 
O
\left(
\begin{array}{c}
\sigma_8\\
\sigma_1
\end{array}
\right) ,\ \ \ \ 
\sigma_n \equiv\frac{u\bar u+d\bar d}{\sqrt{2}},\ \ 
\sigma_s \equiv s\bar s,
\label{eq:or}
\end{eqnarray}
where $O$ is the matrix of the orthogonal transformation, given by 
\begin{eqnarray}
O &\equiv&
\left(
\begin{array}{cc}
 \sqrt{\frac{1}{3}} & \sqrt{\frac{2}{3}} \\
-\sqrt{\frac{2}{3}} & \sqrt{\frac{1}{3}}
\end{array}
\right) .
\label{eq:om}
\end{eqnarray}
Through the transformation $O$,
the elements of the squared-mass matrix in the 
octet-singlet bases are numerically given by
\begin{eqnarray}
\left(
\begin{array}{cc}
 m_{\sigma_8}^2 & m_{\sigma_{81}}^2 \\
 m_{\sigma_{81}}^2 & m_{\sigma_1}^2
\end{array}
\right)  
 &=& 
{}^tO
\left(
\begin{array}{cc}
 m_{\sigma (600)}^2 & 0 \\
 0 & m_{f_0(980)}^2
\end{array}
\right)
O \nonumber\\
 &=&
\left(
\begin{array}{cc}
 (0.87\ {\rm GeV})^2 & -(0.54\ {\rm GeV})^2 \\
 -(0.54\ {\rm GeV})^2 & (0.74\ {\rm GeV})^2
\end{array}
\right) ,
\label{eq:mm}
\end{eqnarray}
where we have used the experimental values $m_\sigma$=0.59 GeV 
and $m_{f_0(980)}$=0.98 GeV.\footnote{
The fact that $m_{\sigma_1}$ is smaller than $m_{\sigma_8}$ 
is in contrast with the case 
of pseudoscalar $\eta$-$\eta '$ mass splitting. 
This fact reflects the property of the 
$U_A(1)$-breaking interaction (see, \S 3).
In this connection, note that the famous nonet mass 
formula\cite{rf:ideal} of the vector mesons,
$m_\rho^2=m_\omega^2$ and $m_\phi^2-m_{K^*}^2=m_{K^*}^2-m_\rho^2$,
is valid in the case $m_{V_1}^2=m_{V_8}^2$.
}

The mass $m_{\sigma_8}$ can also be determined theoretically by using 
the Gell-Mann Okubo (GMO) relation,
\begin{eqnarray}
m_\kappa^2 &=& (3m_{\sigma_8}^{\rm theor\ 2}+m_{a_0}^2)/4 ,
\label{eq:go}
\end{eqnarray}
as 
\begin{eqnarray}
m_{\sigma_8}^{\rm theor} &=& 0.88\ {\rm GeV},
\label{eq:val}
\end{eqnarray}
where we have used the experimental values $m_\kappa$=0.91 GeV 
and $m_{a_0(980)}$=0.98 GeV. This value of 
$m_{\sigma_8}^{\rm theor}$ is quite close 
to $m_{\sigma_8}$=0.87 GeV, obtained
phenomenologically in Eq. (\ref{eq:mm}).
This fact supports our classification that ${\mib \sigma}${\bf (600),} 
${\bf f_0}${\bf (980),}
${\mib \kappa}${\bf (900) and }
${\bf a_0(980)}$ 
{\bf form a single scalar nonet.}  



\section{Chiral symmetry and properties of
the $\sigma$-nonet in relation with the $\pi$-nonet}

\subsection{
Chiral symmetry and the mass and width of the scalar nonet
}

\begin{figure}[t]
 \epsfysize=12.0 cm
 \centerline{\epsffile{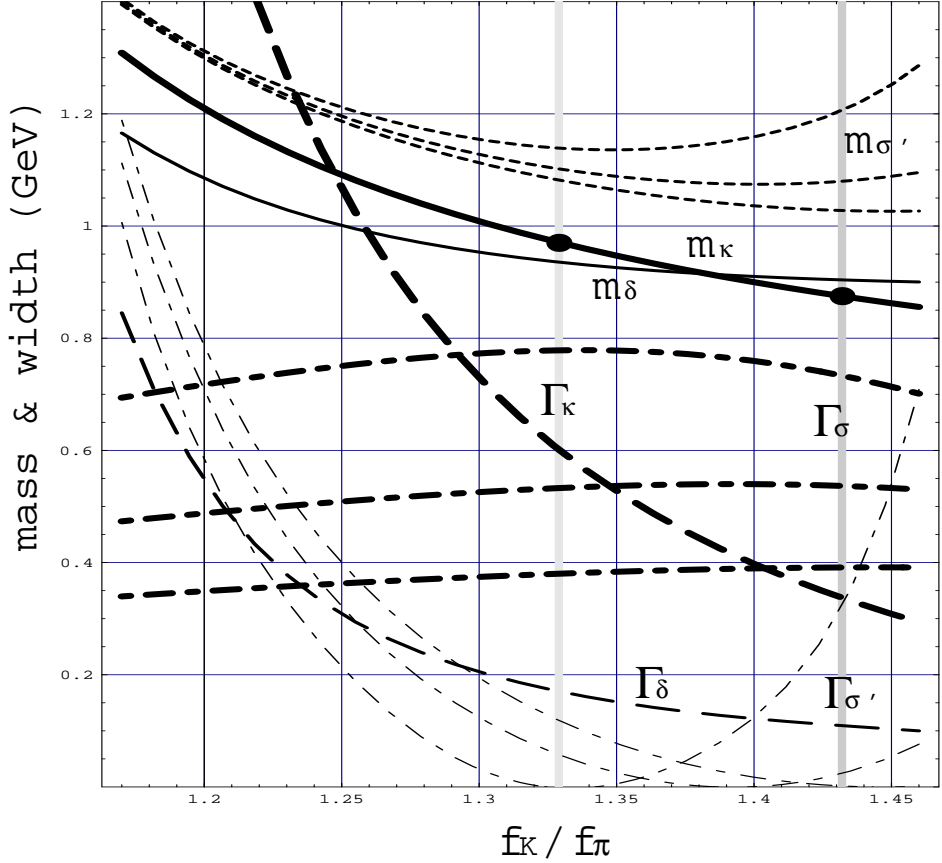}}
 \caption{The scalar meson masses and widths (GeV) 
versus $f_K/f_\pi$.
The upper, middle and lower lines of $M_\sigma '$, $\Gamma_\sigma$
and $\Gamma_{\sigma '}$ 
correspond, respectively, 
to the input 
values $M_\sigma =650,\ 585$ and 535 MeV.
This figure was extended from the original figure drawn by 
Chan and Haymaker,\cite{rf:CH} including the widths of 
scalar mesons. 
}
\label{fig:tree}
\end{figure}

We assume that our scalar $\sigma$-nonet is a composite 
$q\bar q$-system as a chiral partner of the pseudoscalar 
$\pi$-nonet and that 
in the low energy region, where the 
structure of composite mesons can be neglected, 
they can $effectively$ be described by 
the linear $\sigma$ model (L$\sigma$M).
In the matrix 
notation $B\equiv s +i\phi$ ($s\equiv\lambda^i s^i/\sqrt{2}$ and 
$\phi\equiv\lambda^i\phi^i/\sqrt{2}$
denoting the scalar and pseudoscalar meson nonet, respectively),
the Lagrangian 
of $SU(3)$L$\sigma$M\cite{rf:gg,rf:CH} is
\begin{eqnarray}
{\cal L}^{{\rm L}\sigma{\rm M}} &=& 
\frac{1}{2}\langle \partial_\mu B\partial^\mu B
^\dagger 
\rangle 
-\frac{\mu^2}{2}\langle BB
^\dagger 
\rangle
-\frac{\lambda_1}{4}\langle BB^\dagger \rangle^2
-\frac{\lambda_2}{2}\langle (BB^\dagger )^2\rangle
\nonumber\\  
 &+& \kappa_d({\rm det}\ B+{\rm det}\ B^\dagger )
+\langle f s \rangle ,
\label{eq:lsm3}
\end{eqnarray}
where $\langle\ \ \rangle$ represents the trace.
Here $f$ is proportional to the current-quark mass matrix in QCD
and the form  
 $f={\rm diag}\{ f_n,f_n,f_s\} $ guarantees the PCAC.
In the process of spontaneous chiral symmetry breaking,
$s$ acquires the vacuum expectation value 
$s_0\equiv \Sigma ={\rm diag}\{ a,a,b\}$, and 
$s\phi\phi$-couplings appear. The 
pseudoscalar decay constants $f_\pi$ and $f_K$,
and their ratio are represented by 
\begin{eqnarray}
f_\pi &=& \sqrt{2}a,\ 
f_K=\frac{a+b}{\sqrt{2}};\ \ 
\frac{f_K}{f_\pi}
=\frac{a+b}{2a}. 
\label{eq:fp}
\end{eqnarray}
The six model parameters
contained in Eq. (\ref{eq:lsm3}) 
are determined by
the masses of $\pi ,\eta , \eta '$, $\sigma$ and $\kappa$ , 
and the decay constant $f_\pi$, and thus
we can predict the masses and widths
 of the scalar mesons. 
These are given in Table \ref{tab:tree}. 
The predicted properties 
are very sensitive to the value of 
$f_K/f_\pi$,\cite{rf:CH} 
as shown in Fig. \ref{fig:tree}.\cite{rf:CH,rf:NPA}
The deviation of the value of $f_K/f_\pi$
from 1 represents the degree of $SU(3)$ breaking 
by $s_0$, as can be seen from Eq. (\ref{eq:fp}). 
We prefer the region in which  
this ratio satisfies $1.329<f_K/f_\pi<1.432$ (which is somewhat 
larger than the experimental value 1.22)
indicated by the  two vertical lines in the figure,
where the value $m_\kappa^{\rm theor}$ 
reproduces the experimental value 
within its uncertainty.
In this region, $\Gamma_\sigma$ and $\Gamma_\kappa$ 
are obtained with much larger values than those of 
$\Gamma_{\sigma '}$ and $\Gamma_\delta$. 
The reason that $\Gamma_{\kappa \rightarrow K\pi }$,
in spite of its comparatively smaller phase space,
becomes as large as $\Gamma_{\sigma \rightarrow\pi\pi }$
is due to the contribution 
to the coupling constant $g_{\kappa  K\pi}$  
from the determinant-type interaction in Eq. (\ref{eq:lsm3}).

\begin{table}
\begin{center}
\caption{The properties of the scalar meson nonet
predicted by $SU(3)$L$\sigma$M, compared with experiments. 
The underlined values of $m_\sigma$ and $m_\kappa$
along with $f_\pi$, $m_\pi$, 
$m_\eta$ and $m_{\eta '}$ are used as inputs.
The region of the value of $m_\kappa^{\rm exp}$ corresponds to 
the region in which the ratio of the  
decay constants satisfies 
$1.329<f_K^{{\rm L}\sigma{\rm M}}/f_\pi^{{\rm L}\sigma{\rm M}}<1.432$.
The properties of $\delta$ and $\sigma '$
become close to those of the observed resonances
$a_0(980)$ and $f_0(980)$, respectively,
 taken as the experimental candidates. 
The quantity $\Gamma_{\sigma '}^{\rm theor}$ is the partial width
$\Gamma_{\sigma '\rightarrow\pi\pi}^{\rm theor}$.
The value of $\Gamma_{\sigma '\rightarrow KK}^{\rm theor}$
is highly dependent on $m_{\sigma '}^{\rm theor}$,
since $m_{\sigma '}^{\rm theor}$ is close to 
$K\bar K$-threshold.
 }
\begin{tabular}{lcccc}
\hline
\hline
   & $m^{\rm theor}$/MeV & $m^{\rm exp}$/MeV
   & $\Gamma^{\rm theor}$/MeV & $\Gamma^{\rm exp}$/MeV\\
\hline
  $\sigma$ & $\underline{535\sim 650}$ & $\underline{535\sim 650}$
           & $400\sim 800$       & $385\pm 70$ \\
\hline
  $\kappa$ & $\underline{905\stackrel{+65}{\scriptstyle -30}} $
           & $\underline{905\stackrel{+65}{\scriptstyle -30}}$  
   & $300\sim 600$ & $545\stackrel{+235}{\scriptstyle -110} $\\
\hline
  $\delta =a_0(980)$ & $900\sim 930$ & $982.7\pm 2.0$  
           & $110\sim 170$ & $57\pm 11$ \\
\hline
  $\sigma '=f_0(980)$ & $1030\sim 1200$  & $993.2\pm 9.5$
             &  $0\sim 300$   & $67.9\pm 9.4$\\
\hline
\end{tabular}
\label{tab:tree}
\end{center}
\end{table}

The predicted widths of $\sigma$ and $\kappa$
are consistent with the experimental values. 
The predicted masses and widths of the other members,
$\delta (I=1)$ and $\sigma '(s\bar{s})$, are
close to those of
$a_0(980)$ and $f_0(980)$, respectively. 
The $\sigma '$ and $\delta$ states have large $K\bar K$-coupling
constants.\footnote{In the case $f_\pi =93$ MeV, $f_K/f_\pi =1.394$
and $m_\sigma =585$ MeV,
for $\sigma '$, 
${\cal L}_{\rm int}=g_{\sigma '\pi\pi}\sigma '{\mib \pi}^2
+g_{\sigma 'KK}\sigma '(K^+K^-+K^0\bar K^0),
g_{\sigma '\pi\pi}=-0.02\ {\rm GeV},
\ {\rm and}\ g_{\sigma 'KK}=-4.97\ {\rm GeV}$.\\
For $\delta$, 
${\cal L}_{\rm int}=g_{\delta\pi\eta}
(\pi^-\delta^++\pi^+\delta^-+\pi^0\delta^0)\eta
+g_{\delta KK}(\delta^0\frac{K^+K^--K^0\bar K^0}{\sqrt 2}) 
+\delta^+K^0K^-+\delta^-\bar K^0K^+),\ 
g_{\delta\pi\eta}=-3.12\ {\rm GeV},
\ {\rm and}\ g_{\delta KK}=-3.19\ {\rm GeV}.$
}
(Especially the $\sigma '$ strongly couples
to the $K\bar K$ channel.)
This suggests that these states may also  
 be interpreted as 
$K\bar K$-molecule states.\cite{rf:Isg}

In L$\sigma$M, $\sigma $ and $\delta $
have almost the same quark content. Despite this fact,
$m_\delta$ becomes much larger than $m_\sigma$.
This phenomenon is explained by the properties of the  
instanton-induced $U_A(1)$-breaking determinant-type 
interaction.\footnote{
However, see another viewpoint on this problem
 given by Jaffe.\cite{rf:Ja}
}

Thus, it may be plausible to regard 
${\mib \sigma}${\bf (600)}, ${\mib \kappa}${\bf (900)}, 
{\bf a}${}_0${\bf (980)}, {\bf and} {\bf f}${}_0${\bf (980)}
{\bf as members of the scalar nonet,
forming with the members of} ${\mib \pi}${\bf -nonet a linear 
representation of the SU(3) chiral symmetry}.\footnote{
This interesting assignment was suggested
and  insisted upon repeatedly by 
M. D. Scadron.\cite{rf:sca} 
}

All the above results are obtained  in 
the tree level bases of the L$\sigma$M. 
Following the renormalization procedure of  
Chan and Haymaker,\cite{rf:CH} 
we have made a preliminary estimate of the one-loop effects with 
the present renewed experimental data. 
The width at the one-loop level
is defined from the imaginary part of the inverse propagator
of the relevant particle, and it is almost equal to the value
given in Table I.
The masses of the scalar nonet (especially the $m_{\sigma '}$)
is affected by the comparatively large one-loop effect, and 
$m_{\sigma '}^{\rm 1-loop}$ becomes much larger than  
$m_{f_0(980)}$. 
We might expect that introduction of the form factor,
reflecting the composite structure of the relevant meson system,
decreases the effects and leads to the improvement.

Another problem of our scalar assignment is, as mentioned above, 
our theoretical value of $f_K/f_\pi$ is\footnote{
A possible solution may be given by taking into account
the pseudoscalar-axial vector mixing effect, which induces
the field renormalization of pseudoscalars (see 
Ref. \citen{rf:had97b}).
}
somewhat larger than the experimental one.

\subsection{The GMO mass formula, OZI rule and effective 
linear $\sigma$ model}

In \S 2 it is shown that the GMO mass formula 
is approximately satisfied for
the octet members of our $\sigma$-nonet of scalar mesons, where 
$\sigma$(600) and $f_0(980)$ are assumed to be ideal mixing states
of isoscalars $f_8$ and $f_1$.
This assumption was motivated by reasoning 
concerning the decay properties of $f_0(980)$,
based on the approximate validity of the OZI rule.

In \S 3.1, by using the L$\sigma$M {\it at the tree level},
the properties of the $\sigma$-nonet are analyzed quantitatively. 
In this sub-section, based on the results of \S 3.1, 
the validity of the GMO mass formula and the OZI rule in L$\sigma$M 
is examined.

In the L$\sigma$M Eq. (\ref{eq:lsm3}), 
since the explicit symmetry breaking
term $\langle fs\rangle$ is introduced in the $T_3^3$-breaking
pattern, the GMO mass formula is expected to be satisfied.\footnote{
Due to the $T_3^3$-breaking pattern of $f$, $s$ acquires the  
vacuum expectation value $s_0 =\Sigma 
={\rm diag}\{ a,a,b  \}$ of the  $T_3^3$-breaking pattern.
Correpondingly, actual scalar and pseudoscalar mass spectra
include, through the $\lambda_1$, $\lambda_2$ and $\kappa_d$-terms,
higher order effects of the  $T_3^3$-breaking of the $f$ matrix,
and strictly GMO mass formula is expected to be 
valid only approximately in the L$\sigma$M.  
}
On the other hand, the validity of the OZI rule for $s\phi\phi$ 
couplings is generally not guaranteed in the L$\sigma$M.
In Eq. (\ref{eq:lsm3}), 
the two terms with the coefficients 
$\lambda_1$ and $\kappa_d$ give generally OZI-forbidden 
 $s\phi\phi$-couplings.
In the following, among their component couplings,
we consider only the coupling for the octet pseudoscalar
$\phi$-meson (that is, $\langle\phi\rangle =0$),
since only the $\pi\pi$ and $KK$ decay channels
are relevant.
These terms include the following:
\begin{eqnarray}
-2\kappa_d\langle s '\phi^2\rangle
+\kappa_d\langle s '\rangle\langle\phi^2\rangle
-\lambda_1\langle\Sigma s '\rangle\langle\phi^2\rangle .
\label{eq:OZI}
\end{eqnarray}
In Eq. (\ref{eq:OZI}), the second and third terms are 
OZI-forbidden. They are explicitly represented in the ideal bases
($\sigma_n$ and $\sigma_s$), and 
Eq. (\ref{eq:OZI}) is rewritten into the form  
\begin{eqnarray}
-2\kappa_d\langle s '\phi^2\rangle
-\{ \sqrt{2}(\lambda_1a-\kappa_d)\sigma_n
+(\lambda_1b-\kappa_d)\sigma_s\}\langle\phi^2\rangle .
\label{eq:OZI2}
\end{eqnarray}

On the other hand, in the ideal bases, the 
elements of the squared mass matrix of the iso-singlet scalar 
mesons are given by 
\begin{eqnarray}
m^2_{\sigma_n\sigma_n}
  &=& \mu^2+\lambda_1(2a^2+b^2)+4\lambda_1a^2
      +6\lambda_2a^2-2\kappa_db \nonumber\\
m^2_{\sigma_s\sigma_s}
  &=& \mu^2+\lambda_1(2a^2+b^2)+2\lambda_1b^2
      +6\lambda_2b^2 \nonumber\\
m^2_{\sigma_n\sigma_s}
  &=& 2\sqrt{2}a(\lambda_1b-\kappa_d) .
\label{eq:mass}
\end{eqnarray}
In the case of $\lambda_1b=\kappa_d$, 
the value of $m^2_{\sigma_n\sigma_s}$ vanishes and the physical
$\sigma (\sigma ')$ becomes identical to the ideal state
$\sigma_n(\sigma_s)$. 
Moreover, in this case for $\sigma_s$, 
as shown in Eq. (\ref{eq:OZI2}),
the OZI-forbidden second and third terms 
in Eq. (\ref{eq:OZI}) cancel each other, and the 
OZI-allowed first term predicts 
$g_{\sigma_s\pi\pi}=0$, giving the vanishing decay width
$\Gamma_{\sigma '\pi\pi}=0$.
In Fig. 1 this case corresponds to the zero-points on the 
$f_K/f_\pi$-axis of the $\Gamma_{\sigma '\pi\pi}$-curves.
Furthermore, in this case
for $\sigma_n$, the second and third terms approximately 
cancel each other, and their sum becomes small
comparatively to the first term,
and the OZI-rule is almost satisfied. 

The region of the values of $f_K/f_\pi$
chosen in \S 3.1 is close to this case ($\lambda_1b=\kappa_d$), 
and thus the OZI-forbidden term is expected to be small.
For example, in the case that $f_\pi=0.093$ GeV,
$f_K/f_\pi=1.394$ and 
$m_\sigma=0.585$ GeV, the parameters $a,b,\lambda_1$ and $\kappa_d$
are determined, respectively, as 
0.0658 GeV,0.1176 GeV,13.03 and 1.518 GeV.  
The coefficients of the OZI-forbidden terms 
for $\sigma_s$ and $\sigma_n$ in Eq. (\ref{eq:OZI2})
are given, respectively,  by $\lambda_1b-\kappa_d=0.0135$ GeV and 
$\lambda_1a-\kappa_d=-0.662$ GeV. These values are much smaller
than the coefficients $2\kappa_d=3.04$ GeV of the OZI-allowed first 
term in Eq. (\ref{eq:OZI2}),
which may be considered as a typical OZI-allowed coupling
term in L$\sigma$M.

Thus we may conclude that the situation of the approximate validity of
the OZI rule, which was the basis of our assignment of 
$\sigma$ and $f_0(980)$ in \S 2, is actually realized in SU(3)
L$\sigma$M with our classified members. 


\section{Concluding Remarks}

In this paper we have investigated the possibility of  
classification of the new scalar nonet, 
$\sigma$(600), $\kappa$(900), $a_0(980)$ and $f_0(980)$.
First by assuming the approximate validity of the OZI rule 
for the decay property of $f_0(980)$, 
the $\sigma$(600) and  $f_0(980)$ were supposed to 
be ideal mixing states, $n\bar n$ and $s\bar s$, respectively,
of the scalar $\sigma$-nonet. It was then pointed out 
that the mass value of the iso-singlet flavor-octet state,
obtained from the orthogonal transformation, satisfies the 
Gell-Mann Okubo mass formula. 
Furthermore, it was shown that 
the experimental masses and widths of members of 
this scalar $\sigma$-nonet are 
consistent with those of the scalar nonet 
predicted by (the tree level calculation of) the L$\sigma$M,
as shown in Table I. 

This result implies that the chiral symmetry plays 
a stronger role than ever thought in understanding 
the strong interaction, not only for deriving the low 
energy theorems through the non-linear 
realization,
but also for explaining the spectroscopy and reactions related to 
 all the mesons with masses below and around $\sim$ 1 GeV
through the linear realization.\\

We now give 
supplementary discussions related to the present problem. 
It is often argued that
the validity of the L$\sigma$M and the 
existence of the $\sigma$-meson {\it as a chiral partner 
of the $\pi$-meson } are not acceptable, 
since the phenomenological pattern of the 
ten low energy coefficients of the $O(p^4)$-level
of chiral perturbation theory
(ChPT)\cite{rf:basic,rf:GL,rf:skir,rf:pich,rf:don}
is not reproduced by the L$\sigma$M.
Surely, the framework of ChPT
is useful for  relating phenomenologically the various
{\em low energy} phenomena concerning the
Nambu-Goldstone $\pi$-meson octet mutually, with the ten parameters.
However, the above argument seems too excessive and  
not appropriate, since
it is  based only on the results of analyses of indirect 
experiments  
with much lower energy
than $m_\sigma$. 
Whether or not a resonance exists should be investigated directly 
by experiments with sufficiently 
high energy to produce the relevant resonance.
As a matter of fact,  $\sigma$(600) can now be  
directly observed both in $\pi\pi$ scattering and in production 
processes.\cite{rf:shim,rf:GAMS,rf:taku1,rf:taku2,rf:JPsi}
\cite{rf:had97,rf:had97d,rf:had97c,rf:CB}
Furthermore,
the parameters in the L$\sigma$M  
describe the physics in the  
{\em resonance energy region as well as the low energy region},
and the contribution of the $\sigma$-meson to low energy quantities 
can be predicted with no $new$ free parameters.\footnote{
In this connection we refer the reader to 
an interesting work which points out that
the width and form factor of $K_{l4}$-decay,
as well as the $\pi\pi$-phase shift $\delta_0^0$
in low energy region,
are also approximately described by $SU(3)$L$\sigma$M,\cite{rf:shaba}
as in the case of chiral perturbation theory.
}

On the other hand
it has been discussed\cite{rf:meissner} that 
in the framework of ChPT
the effect of the $\sigma$-meson can be taken into account through
the $O(p^4)$ and $O(p^6)$ Lagrangian.
However, such an approach 
seems clearly to have no predictive power
regarding the properties of the 
 scalar $\sigma$-meson nonet itself
{\it as a chiral partner of the $\pi$-nonet.}

Finally we would like to mention that, as we have pointed out 
previously,\cite{rf:pipip,rf:NPA}
the $\sigma$-nonet treated in this paper should be 
discriminated from the $^3P_0$-nonet.
The $\sigma$-nonet is  assigned the 
quantum numbers $(L,S)=(0,0)$ 
in the ``relativistic $LS$-coupling scheme.''
 

\section*{Acknowledgements}
\addcontentsline{toc}{section}
{Acknowledgements}

I am grateful to Professor M. Oka for his continual
encouragement. 
The contents of the present work are closely connected with a 
series of works done in cooperation 
with the theoretical physics group of Nihon University.
I would like to thank Professor S. Ishida
 for many useful comments, and
I would like to express my sincere appreciation to 
all members of the NU-group.
I would also like to thank Professor K. Takamatsu, Professor T. Tsuru
and Doctor T. Ishida for suggesting many ideas. 
I am grateful to Professor K. Yazaki, Professor S. Wada,
Professor J. Arafune, Professor K.Ohta and Professor T. Kobayashi
for their valuable comments, especially on the mass formula given 
in \S 2.


\end{document}